\documentclass[a4paper,conference]{IEEEtran}
\ifCLASSINFOpdf
\else
\fi
\usepackage{amsmath,amsthm,amssymb}
\usepackage{breqn}
\usepackage{multirow}
\usepackage{subfig}
\usepackage{graphicx}
\begin{document}
%
\title{Adversarially Training for Audio Classifiers}

\author{\IEEEauthorblockN{Raymel Alfonso Sallo}
\IEEEauthorblockA{\'{E}cole de Technologie Sup\'{e}rieure (\'{E}TS)\\
D\'{e}partement de G\'{e}nie Logiciel et TI\\
1100 Notre-Dame W, Montr\'{e}al,\\ H3C 1K3, Qu\'{e}bec, Canada\\
Raymel.Alfonso-Sallo.1@ens.etsmtl.ca}
\and
\IEEEauthorblockN{Mohammad Esmaeilpour}
\IEEEauthorblockA{\'{E}cole de Technologie Sup\'{e}rieure (\'{E}TS)\\
D\'{e}partement de G\'{e}nie Logiciel et TI\\
1100 Notre-Dame W, Montr\'{e}al,\\ H3C 1K3, Qu\'{e}bec, Canada\\
Mohammad.Esmaeilpour.1@ens.etsmtl.ca}
\and
\IEEEauthorblockN{Patrick Cardinal}
\IEEEauthorblockA{\'{E}cole de Technologie Sup\'{e}rieure (\'{E}TS)\\
D\'{e}partement de G\'{e}nie Logiciel et TI\\
1100 Notre-Dame W, Montr\'{e}al,\\ H3C 1K3, Qu\'{e}bec, Canada\\
Patrick.Cardinal@etsmtl.ca}}


%


\maketitle

\begin{abstract}
In this paper, we investigate the potential effect of the adversarially training on the robustness of six advanced deep neural networks against a variety of targeted and non-targeted adversarial attacks. We firstly show that, the ResNet-56 model trained on the 2D representation of the discrete wavelet transform appended with the tonnetz chromagram outperforms other models in terms of recognition accuracy. Then we demonstrate the positive impact of adversarially training on this model as well as other deep architectures against six types of attack algorithms (white and black-box) with the cost of the reduced recognition accuracy and limited adversarial perturbation. We run our experiments on two benchmarking environmental sound datasets and show that without any imposed limitations on the budget allocations for the adversary, the fooling rate of the adversarially trained models can exceed 90\%. In other words, adversarial attacks exist in any scales, but they might require higher adversarial perturbations compared to non-adversarially trained models.  
\end{abstract}


%
\IEEEpeerreviewmaketitle

\section{Introduction}
The existence of adversarial attacks has been characterized for data-driven audio and speech recognition models for both waveform and representation domains \cite{carlini2018audio, esmaeilpour2019robust}. During the past years, many strong white and black-box adversarial algorithms have been introduced which they basically recast costly optimization problems against victim classifiers. Unfortunately, these attacks effectively degrade the classification performance of almost all data-driven models from conventional classifiers (e.g., support vector machines) to the state-of-the-art deep neural networks \cite{esmaeilpour2020detection}. This poses an extreme growing concern about the security and the reliability of the classifiers.

The typical approach in crafting adversarial example is to solve an optimization problem in order to obtain the smallest possible perturbations for the legitimate samples, undetectable by a human, aiming at fooling the classifier. The commonly used measures to compare the altered sample with the original one are $l_{2}$ or $l_{\infty}$ similarity metrics. The computational complexity of this optimization process is dependent to the dimensions of the given input samples. Consequently, it requires considerable computational overhead for high dimensional data, even in the case of short audio signals \cite{carlini2018audio}. However, regardless of the computational cost of the attacks, this threat actively exists for any end-to-end audio and speech classifier. Since the highest recognition accuracies have been reported on 2D representations of audio signals \cite{esmaeilpour2019robust, boddapati2017classifying}, the optimized attack algorithms developed for computer vision applications such as fast gradient sign method (FGSM) \cite{goodfellow2014explaining} led to security concerns for audio classifiers \cite{esmaeilpour2020detection}.

Although some approaches have been introduced for defending victim models against adversarial attacks, there is not yet a reliable framework achieving the required efficiency. Based on the detailed discussion in \cite{athalye2018obfuscated}, common defence algorithms usually obfuscate gradient information but running stronger attack algorithms against them consistently fool these detectors. Unfortunately, vulnerability against adversarial attacks is an open problem in data-driven classification and though the generated fake examples look very similar to noisy samples, they lie in dissimilar subspaces \cite{esmaeilpour2020detection, ma2018characterizing}. It has been shown that adversarial examples lie in the manifolds marginally over the decision boundary of the victim classifier, where the model lacks of generalizability \cite{esmaeilpour2020detection}. Therefore, integrating these examples into the training set of the victim classifier could improve the robustness. This approach, known as adversarially training \cite{goodfellow2014explaining}, might be a more reasonable defense approach without shattering gradient vectors \cite{athalye2018obfuscated}. However, there is no guarantee for the safety of the adversarially trained classifiers \cite{tramer2017ensemble}.

Although there are some discussions in the computer vision domain about the negative effect of adversarially training on the recognition performance of the victim classifiers \cite{papernot2016transferability}, to the best of our knowledge, this potential side effect has not been yet studied for the 2D representation of audio signals. We address this issue in this paper and report our results on two benchmarking environmental sound datasets. Specifically, our main contributions in this paper are:
\begin{itemize}
    \item characterizing the adversarially training impact on six advanced deep neural network architectures for diverse audio representations,  
    \item demonstrating that deep neural networks specially those with residual blocks have higher recognition performance on tonnetz features concatenated with DWT spectrograms compared to STFT representations,
    \item showing the adversarially trained AlexNet model outperforms ResNets with limiting the perturbation magnitude,
    \item experimentally proving that although adversarially training reduces recognition accuracy of the victim model, it makes the attack more costly for the adversary in terms of required perturbation.
\end{itemize}

The rest of this paper is organized as follows. In Section~\ref{sec:relatedWorks}, we review some related works about adversarial attacks developed for 2D domains. Details about signal transformation and 2D representation production are provided in Section~\ref{sec:audiorep} and \ref{sec:specgen}, respectively. In Section~\ref{sec:modelsg}, we briefly introduce our selected front-end audio classifiers which are state-of-the-art deep learning architectures. The adversarial attack procedures and budget allocation for the adversary are discussed in Section~\ref{adv:setup}. Accordingly, section~\ref{sec:advtra} explains the adversarially training framework and obtained results are summarized in Section~\ref{sec:expersut}. 

\section{Related Works}
\label{sec:relatedWorks}
There is a large volume of published studies on attacking classifiers using different optimization techniques aiming to effectively disrupt their recognition performances. In this paper, we focus on five strong white-box targeted and non-targeted attack algorithms which have been reported to be very destructive when used on advanced deep learning models trained on audio representations \cite{esmaeilpour2019robust}. Moreover, we also use a black-box adversarial attack, based on the gradient approximation, against the victim classifiers . 

The fast gradient sign method is a well-established baseline in targeted adversarial attack. The computational cost of this one-shot approach at runtime is low, taking advantage of the linear characteristics in deep neural networks. Kurakin \textit{et al.} \cite{kurakin2016adversarial} introduced an iterative version of FGSM, known as the basic iterative method (BIM), for running stronger attacks against victim classifiers and is formulated at:
\begin{equation}
\bold{x{}'}_{n+1}=\mathrm{clip}\left \{ {\bold{x{}'}_{n}} +\zeta  \mathrm{sgn}\left ( \nabla_{\bold{x}
} J\left [ \bold{x{}'}_{n},l(\bold{x}) \right ] \right ) \right \}
\label{eq:bimab}
\end{equation}
\noindent where the legitimate and its associated adversarial examples are represented by $\bold{x}$ and $\bold{x{}'}$, respectively. The initial state in this recursive formulation is $\bold{x{}'}_{0} = \bold{x}$ in the $\epsilon$-neighbourhood (the distance measured by a similarity metric such as $l_{2}$) of the legitimate manifold. This is followed by a clipping operation for keeping the adversarial perturbation within $\left [ -\epsilon, \epsilon \right ]$. Moreover, $l(\bold{x})$ and $\mathrm{sgn}(\cdot)$ stand for the label of $\bold{x}$ and the general sign function. In Eq.~\ref{eq:bimab}, the step size $\zeta=1$, though it is tunable according to the adversary's wishes. Two types of optimizations can be used with Eq.~\ref{eq:bimab}: (1) optimizing up to reach the first adversarial example (BIM-a) and (2) continuing the optimization up to a predefined number of iterations (BIM-b). For measuring the $\epsilon$, two similarity metrics are suggested: $l_{\infty}$ and $l_{2}$. In this work, we focus on the latter. 

Gradient information of a deep neural network contains direction of intensity variation associated with the model decision boundary. Exploiting these information vectors for finding the least likely probability distribution is the key idea of the Jacobian-based Saliency map attack (JSMA) \cite{papernot2016limitations}. For the adversarial label $l{}'$, this iterative attack algorithm runs against the model $f$ and strives to achieve $l{}'=\arg \max_{j} f_{j}(\bold{x})$. The JSMA increases the probability of the target label $l{}'$ while minimizes those of the other classes including the ground-truth using a saliency map as shown in Eq.~\ref{eq:jsma}.
\begin{equation}
    S(\bold{x}, l{}')[i]=\left | J_{i,l{}'}(\bold{x}) \right |\left ( \sum_{j\neq l{}'}J_{i,j}(\bold{x}) \right )
    \label{eq:jsma}
\end{equation}
\noindent where $J_{i,j}$ denotes the forward derivative of the model for the feature $i$ computed as:
\begin{equation}
    J_{f}[i,j](\bold{x})=\frac{\partial f_{j}(\bold{x})}{\partial \bold{x}_{i}}
    \label{eq:jsma2}
\end{equation}
\noindent the Jacobian vectors associated with label $l{}'$ and values of the saliency map less or greater than zero (no variation shield), $ S(\bold{x}, l{}')[i]=0$. This white-box attack algorithm searches, iteratively, the feature index on which the perturbation will be applied in order to fool the model toward the target label $l{}'$ using the similarity metric $l_{0}$.

The perturbation required for pushing a sample over the decision boundary of the victim classifier should be as minimal as possible. In a white box scenario, the optimization process uses local properties of the decision boundary. It has been shown that linearizing the boundary in the subspace of the original samples can yield to adversarial perturbation smaller than FGSM attack. This approach, known as the DeepFool attack, is shown in Eq.~\ref{eq:deepfool} \cite{moosavi2016deepfool}:
\begin{equation}
 \arg \min \left \| \epsilon \right \|_{2}, \quad \epsilon = -f(\bold{x})\bold{w}/\left \| \bold{w} \right \|_{2}^{2}
 \label{eq:deepfool}
\end{equation}
\noindent where the $\bold{w}$ refers to the weight function of the recognition model. Unlike other abovementioned adversarial attacks, DeepFool is a non-targeted attack and it iterates as many times as needed for pushing random samples to be marginally over the locally linearized decision boundary with the condition of maximizing the prediction likelihood toward any labels other than the ground-truth. Though both $l_{\infty}$ or $l_{2}$ measurement metrics can be used in the DeepFool attack, we use the latter in accordance with BIM algorithms.

Presumably, a straightforward approach for keeping an adversarial perturbation undetectable can be achieved by reducing its magnitude and distribute it over all input features. Additionally, not every feature should be perturbed and their gradient vectors should not be shattered. Following these two assumptions, Carlini and Wagner attack (CWA) has been introduced \cite{carlini2017towards}. The general framework of their proposed algorithm is based on the following minimization problem:
\begin{equation}
    \min \left \| \bold{x{}'} - \bold{x} \right \|_{2}^{2} +c \cdot  \mathcal{L}(\bold{x{}'})
    \label{eq:cwa}
\end{equation}
\noindent where the constant $c$ is obtainable through a binary search. Finding the most appropriate value for this hyperparameter is very challenging since it may easily dominate the distance function and push the sample too far away from the adversarial subspace. Although in Eq.~\ref{eq:cwa} the $l_{2}$ similarity metric for computing the adversarial perturbation $\epsilon$ is employed, CWA properly generalizes for both $l_{0}$ and $l_{\infty}$. In the configuration of this adversarial attack, the loss function $\mathcal{L}$ is defined over the logits of $Z$ for the trained model $f$ as shown in the following equation:
\begin{equation}
 \mathcal{L}(\bold{x}{}')=\max \left [ \max_{i\neq l^\prime}\left \{ Z(\bold{x}^\prime)_{i} - Z(\bold{x}^\prime)_{l^\prime}, -\kappa \right \} \right ]
 \label{cwa_func}
\end{equation}
\noindent where $\kappa$ controls the effectiveness and the adjacency of the adversarial examples to the decision boundary of the model. In this regard, higher values for this parameter in conjunction with a minimum $\epsilon$-neighbourhood results in adversarial examples with higher confidence. 

For achieving the overall unrestricted adversarial perturbation ($\left \| \epsilon \right \|_{2}$) with small enough magnitude, CWA solves Eq.~\ref{eq:cwa} through the following optimization framework:
\begin{equation}
    \min_{\rho}\left \| \frac{1}{2}\left ( \tanh(\rho) +1 \right ) - \bold{x} \right \|_{2}^{2} + c\cdot \mathcal{L}\left ( \frac{1}{2}\tanh(\rho)+1 \right )
    \label{eq:cwa2}
\end{equation}
\noindent where $\rho = \arctan(\bold{x}+\delta)$ and the unrestricted approximate perturbation $\delta^{*}$ is as the following.
\begin{equation}
    \delta^{*}=\frac{1}{2}\left ( \tanh(\rho+1) \right )-\bold{x}
\end{equation}
\noindent This perturbation is unrestricted and it should be tuned for feature values by measuring $\nabla f(\bold{x}+\delta^{*})$. For feature intensities with negligible gradient values, the actual adversarial perturbation truncates to zero, and for the rest: $\delta \leftarrow  \delta^{*}$.

Attacking victim classifiers while there is an unrestricted access to the details of the attacked models, including the training dataset, hyperparameters, architecture, and more importantly gradient information, like all the abovementioned attack algorithms, is less challenging compared to the black-box attack scenarios. Usually, in the latter scheme, the adversary runs gradient estimation via querying the classifier by training a surrogate model. In this paper, the chosen black-box attack is the natural evolution strategy (NES \cite{wierstra2008natural}) which has been employed for gradient approximation in \cite{ilyas2018black}. This iterative algorithm is known as partial information attack (PIA) and it encodes $l_{\infty}$ similarity metric as part of its targeted optimization problem. Finding the proper adversarial perturbation bound for PIA is to some extent challenging and requires a very high number of querying to the victim model.


Before discussing how adversarial attack and adversarially training on various deep neural network architectures have been implemented, we firstly need to provide a brief overview on the transformation of an audio signal into 2D representations. The next section will describe spectrogram generation using short time Fourier transformation (STFT), discrete wavelet transformation (DWT), and tonnetz feature extraction. We will then train our classifiers using these representations and investigate how adversarially training impacts their robustness to adversarial attacks.

\section{Audio Transformation}
\label{sec:audiorep}
Since audio and speech signals have high dimensionality in time domain, their 2D representations with lower dimensionalities have been widely used for training advanced classifiers originally developed for 2D computer vision applications \cite{esmaeilpour2020unsupervised}. In this work, we use STFT and DWT, both with and without tonnetz features for generating 2D representations of audio signals. This section briefly reviews the required transformations by this work.

For a discrete signal $a[n]$ distributed over time $n$ using the Hann window function $H[\cdot]$, we can compute the complex Fourier transformation using the following equation:
\begin{equation}
 \mathrm{STFT}\begin{Bmatrix} a[n] \end{Bmatrix}(m,\omega)=\sum_{n=-\infty}^{\infty}a[n]H[n-m]e^{-j\omega n}
 \label{eq:STFT}
\end{equation}
\noindent where $m$ is the time scale and  $m\ll n$. Additionally, $\omega$ stands for the continuous frequency coefficient. This transformation applies on shorter overlapping sub-signals with a predefined sampling rate and forms the STFT spectrogram as shown in Eq.~\ref{eq:STFT_spec}.
\begin{dmath}
 \mathrm{SP_{STFT}}\begin{Bmatrix} a[n] \end{Bmatrix}(m,\omega)=
 \left | \sum_{n=-\infty}^{\infty}a[n]w[n-m]e^{-j\omega n} \right |^2
 \label{eq:STFT_spec}
\end{dmath}

\noindent There are several variants of the STFT transformation such as mel-scale and cepstral coefficient, producing even lower dimensionality, that have been widely used for various speech processing tasks \cite{patel2010speech, juvela2018speech}. However in this work, we use the standard STFT representation for training the front-end dense classifiers.

Generating DWT spectrogram is very similar to the Fourier transformation as they both employ continuous and differentiable basis functions. For the wavelet transformation, several functions have been studied and their effectiveness for audio signals have been investigated in \cite{mitra2008content, patidar2014classification}. The general form of this transformation for a continuous function $a(t)$ is shown in Eq.~\ref{eq:dwt_contin}. 
\begin{equation}
 \mathrm{DWT}\begin{Bmatrix} a(t) \end{Bmatrix} = \frac{1}{\sqrt{\left | s \right |}}\int_{-\infty}^{\infty}a(t)\psi \begin{pmatrix} \frac{t-\tau}{s} \end{pmatrix}dt
 \label{eq:dwt_contin}
\end{equation}
\noindent where $\tau$ and $s$ refer to the time variations in the transformation and the wavelet scale, respectively. Moreover, $\nu $ stands for the basis mother functions. Common choices for this function are Haar, Mexican Hat, and complex Morlet. The latter has been extensively used in signal processing, mainly because of its nonlinear characteristics \cite{esmaeilpour2020unsupervised} (see Eq.~\ref{eq:morlet_func}).
\begin{equation}
 \psi(t)=\frac{1}{\sqrt{2\pi}}e^{-j\omega t}e^{-t^{2}/2}
 \label{eq:morlet_func}
\end{equation}
\noindent The complex Morlet is continuous in its conjugate manifold. The convolution of this function with overlapping chunks of the given audio signal results in its spectral visualization as described in Eq.~\ref{eq:dwt_spectrogram}.
\begin{equation}
 \mathrm{SP_{DWT}}\begin{Bmatrix} a(t) \end{Bmatrix}=\left | \mathrm{DWT}\begin{Bmatrix} a[k,n] \end{Bmatrix} \right |
 \label{eq:dwt_spectrogram}
\end{equation}
\noindent where $k$ and $n$ are integer numbers associated with scales of $\psi$.

The two aforementioned transformations represent spatiotemporal modulation features of a signal in the frequency domain, regardless of its harmonic characteristics. It has been demonstrated that using harmonic change detection function (HCDF) provides distinctive features for the audio signal \cite{harte2006detecting}. This function provides chromagram from the constant-Q transformation (CQT) which are also known as tonnetz features. According to \cite{harte2006detecting}, there are four major steps in a HCDF system. Firstly, the audio signal is converted into a logarithmic spectrum vectors using CQT. Then, pitch-class vectors are extracted from the tonal transformation based on the quantized chromagram. In the third step, 6-dimensional centroid vectors form a tensor from the tonal transformation. Finally, a smoothing operation postprocesses this tensor for distance calculation. 

We use HCDF system for generating spectrogram from audio signals in order to enhance recognition performance of the classifiers. In the next section, we provide details of this process for two benchmarking environmental sound datasets.

\captionsetup[subfigure]{labelformat=empty}
\begin{figure*}[htb]%
 \footnotesize 
 \setlength{\tabcolsep}{23pt}
\centering
 \subfloat[DWT]{{\includegraphics[width=0.135\textwidth]{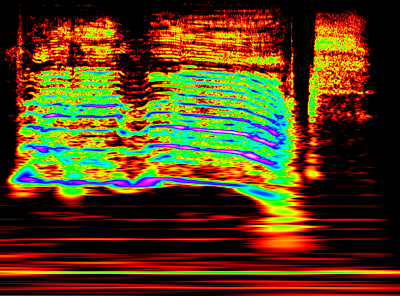} }}
 \subfloat[$\left \| \epsilon \right \|_{2}=1.32$]{{\includegraphics[width=0.135\textwidth]{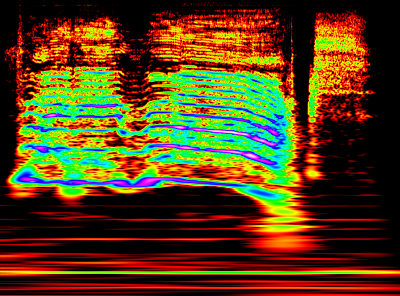} }}
 \subfloat[$\left \| \epsilon \right \|_{2}=1.29$]{{\includegraphics[width=0.135\textwidth]{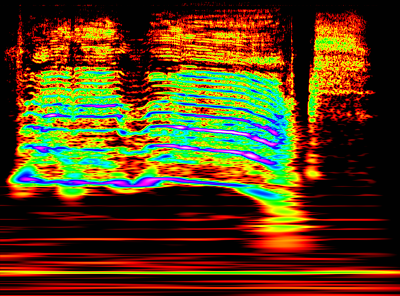} }}%
 \subfloat[$\left \| \epsilon \right \|_{0}=1.07$]{{\includegraphics[width=0.135\textwidth]{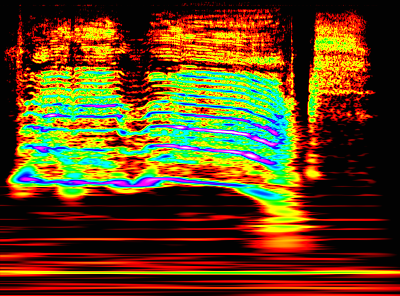} }}%
 \subfloat[$\left \| \epsilon \right \|_{2}=0.49$]{{\includegraphics[width=0.135\textwidth]{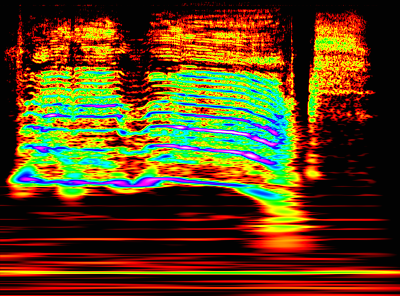} }}%
 \subfloat[$\left \| \epsilon \right \|_{2}=2.18$]{{\includegraphics[width=0.135\textwidth]{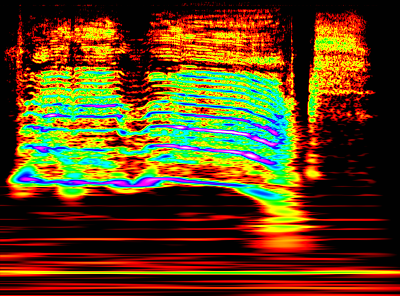} }}%
 \subfloat[$\left \| \epsilon \right \|_{\infty}=1.76$]{{\includegraphics[width=0.135\textwidth]{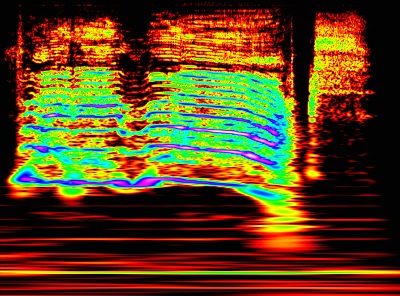} }}%
 \\
 \subfloat[DWT $|$ Tonnetz]{{\includegraphics[width=0.135\textwidth]{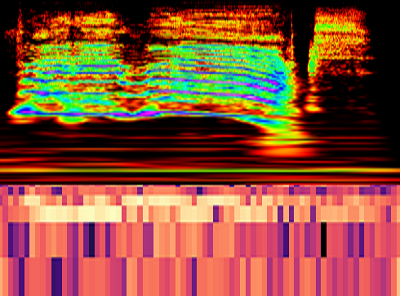} }}%
 \subfloat[$\left \| \epsilon \right \|_{2}=1.93$]{{\includegraphics[width=0.135\textwidth]{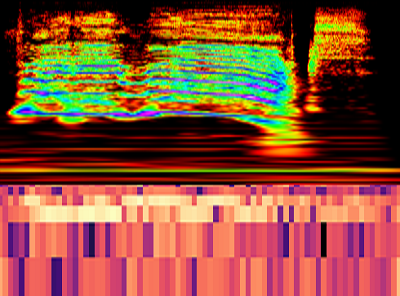} }}%
 \subfloat[$\left \| \epsilon \right \|_{2}=1.21$]{{\includegraphics[width=0.135\textwidth]{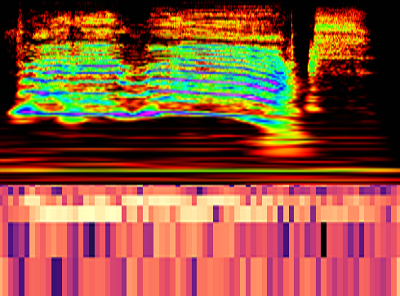} }}%
 \subfloat[$\left \| \epsilon \right \|_{0}=1.59$]{{\includegraphics[width=0.135\textwidth]{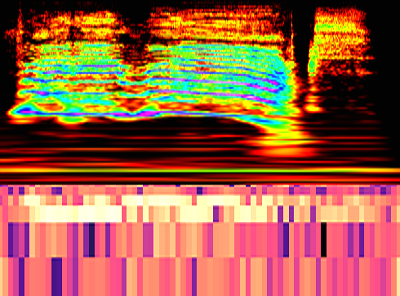} }}%
 \subfloat[$\left \| \epsilon \right \|_{2}=0.91$]{{\includegraphics[width=0.135\textwidth]{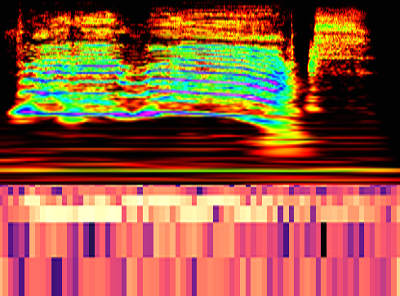} }}%
 \subfloat[$\left \| \epsilon \right \|_{2}=2.69$]{{\includegraphics[width=0.135\textwidth]{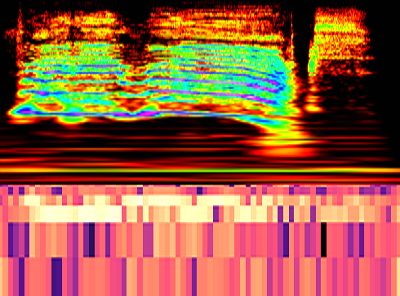} }}%
 \subfloat[$\left \| \epsilon \right \|_{\infty}=1.35$]{{\includegraphics[width=0.135\textwidth]{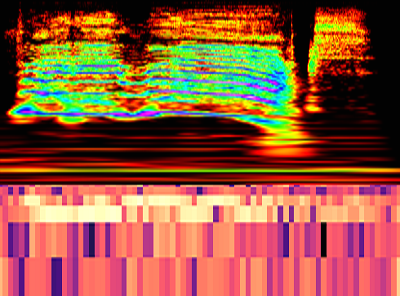} }}%
 \\
 \subfloat[STFT]{{\includegraphics[width=0.135\textwidth]{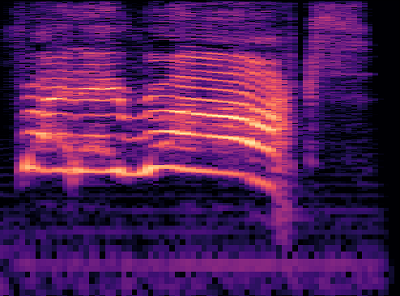} }}%
 \subfloat[$\left \| \epsilon \right \|_{2}=2.15$]{{\includegraphics[width=0.135\textwidth]{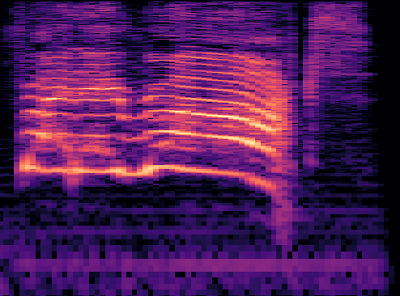} }}%
 \subfloat[$\left \| \epsilon \right \|_{2}=1.18$]{{\includegraphics[width=0.135\textwidth]{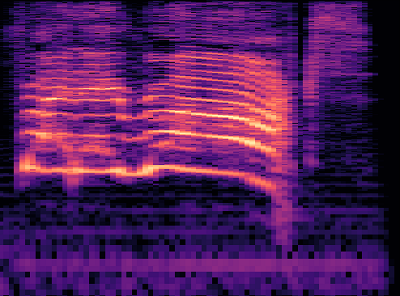} }}%
 \subfloat[$\left \| \epsilon \right \|_{0}=2.28$]{{\includegraphics[width=0.135\textwidth]{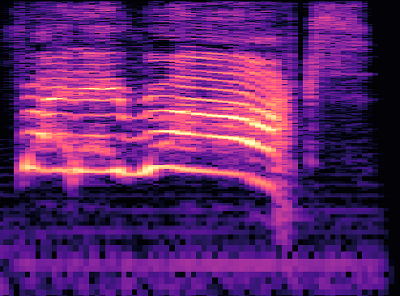} }}%
 \subfloat[$\left \| \epsilon \right \|_{2}=1.98$]{{\includegraphics[width=0.135\textwidth]{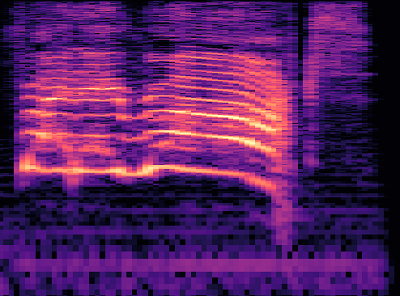} }}%
 \subfloat[$\left \| \epsilon \right \|_{2}=1.83$]{{\includegraphics[width=0.135\textwidth]{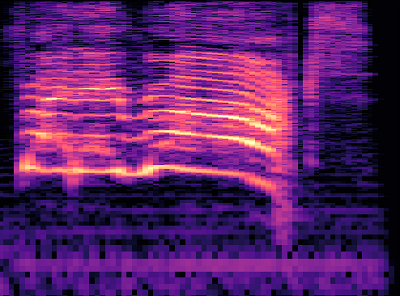} }}%
 \subfloat[$\left \| \epsilon \right \|_{\infty}=2.14$]{{\includegraphics[width=0.135\textwidth]{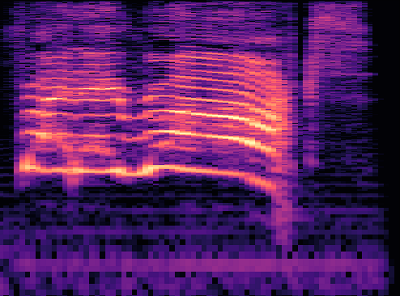} }}%
 \\
 \subfloat[STFT $|$ Tonnetz]{{\includegraphics[width=0.135\textwidth]{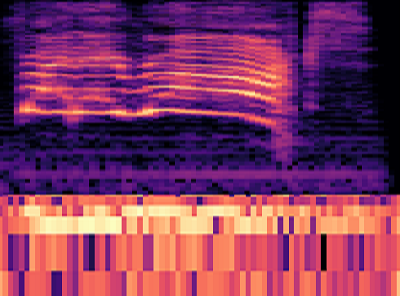} }}%
 \subfloat[$\left \| \epsilon \right \|_{2}=0.84$]{{\includegraphics[width=0.135\textwidth]{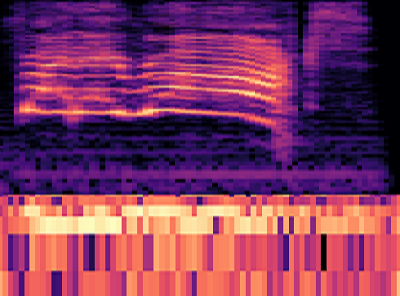} }}%
 \subfloat[$\left \| \epsilon \right \|_{2}=1.63$]{{\includegraphics[width=0.135\textwidth]{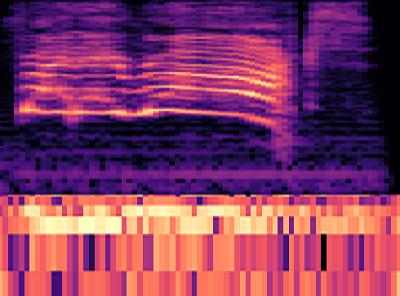} }}%
 \subfloat[$\left \| \epsilon \right \|_{0}=2.51$]{{\includegraphics[width=0.135\textwidth]{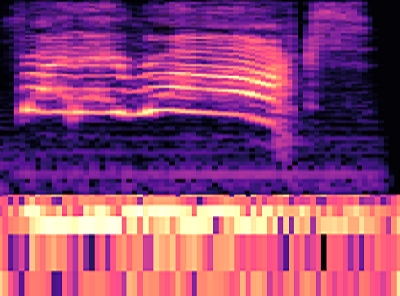} }}%
 \subfloat[$\left \| \epsilon \right \|_{2}=2.65$]{{\includegraphics[width=0.135\textwidth]{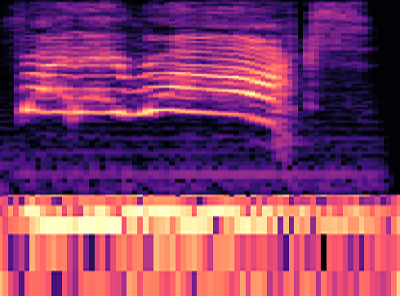} }}%
 \subfloat[$\left \| \epsilon \right \|_{2}=1.88$]{{\includegraphics[width=0.135\textwidth]{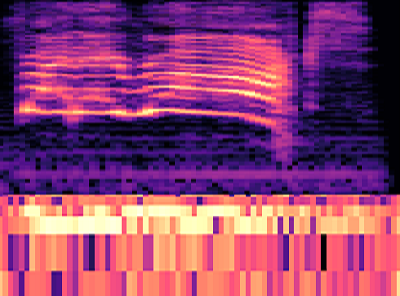} }}%
 \subfloat[$\left \| \epsilon \right \|_{\infty}=2.06$]{{\includegraphics[width=0.135\textwidth]{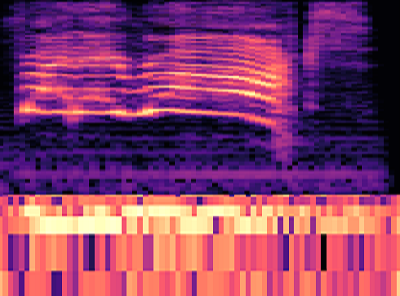} }}%
\caption{Crafted adversarial examples for the ResNet-56 using the six optimization-based attack algorithms. The first column of the figure denotes the original representations for the randomly selected sample from the class of 'children playing' in the UrbanSound8K dataset. Other columns are associated with the attack algorithms namely, BIM-a, BIM-b, JSMA, DeepFool, CWA, and PIA, respectively. Adversarial Perturbation values have been written at the bottom of each adversarial spectrogram.}
\label{attack_spect}
\end{figure*}

\section{Spectrogram Production}
\label{sec:specgen}
We produce STFT representation based on the instructions provided by the open source Python library Librosa \cite{mcfee2015librosa}. We set the windows size and the hop length ($n$ and $m$ in Eq.~\ref{eq:STFT}) to 2048 and 512, respectively. Additionally, we initialize the number of filters to 2048 which is the standard value for the environmental sounds task \cite{esmaeilpour2020unsupervised}. Audio chunks associated with each window are padded in order to reduce the potential negative effect of loosing temporal dependencies. Furthermore, the frames are overlapped using a ratio of 50\%. 

For generating DWT spectrograms, we use our modified version of the wavelet sound explorer \cite{waveletSoundExplorer} with the complex Morlet mother function. As proposed by \cite{boddapati2017classifying}, we set the DWT sampling frequency to 16 KHz for ESC-50 and 8 KHz for UrbanSound8K with the uniform 50\% overlapping ratio. For enhancement purposes, we use the logarithmic visualization on the generated spectrograms to better characterize high frequency areas. 

For the tonnetz chromagram, we use the default settings provided by Librosa with the sampling rate of 22.05 KHz. We resize the resulting chromagrams in such a way that the result will comply with the aforementioned representations. Inspired from \cite{su2019environment}, we append these features to the STFT and DWT spectrograms and organize them into two additional representations. In the next section, we provide more details about the training of the front-end classifiers using these four spectrogram sets.

\section{Classification Models}
\label{sec:modelsg}
Since an adversary runs the adversarial attack against the classifier, the choice of the victim network architecture affects the fooling rate of the model. This issue has been studied in \cite{esmaeilpour2019robust} for the advanced GoogLeNet \cite{szegedy2015going} and AlexNet \cite{krizhevsky2012imagenet} architectures trained on DWT (with linear, logarithmic, and logarithmic real visualizations), STFT, and their pooled spectrograms. Since our main objective is investigating the impact of adversarially training on advanced deep learning classifiers, we additionally include ResNet-X architectures with $X \in \left \{ 18, 34, 56 \right \}$ \cite{he2016deep} and VGG-16 \cite{simonyan2014very} architectures. 

The pretrained models of these six classifiers have been used and the input and output layers have been fine-tuned as described in \cite{esmaeilpour2019robust}. Computational hardware used for all experiments are two NVIDIA GTX-$1080$-Ti with $4 \times 11$ GB memory in addition to a $64$-bit Intel Core-i$7$-$7700$ ($3.6$ GHz) CPU with $64$ GB RAM. We carry out our experiments using the five-fold cross validation setup for all the spectrogram sets. As a common practice in model performance analysis, we preserve 70\% of the entire samples for training and development followed by running the early stopping scenario. We report recognition accuracy of these models for the remaining 30\% samples.

In the next section, we provide the detailed setup for the adversarial algorithms mentioned in section~\ref{sec:relatedWorks}. We additionally discuss budget allocations required by the adversary for successfully attacking the six finely trained victim models.

\section{Adversarial Attack Setup}
\label{adv:setup}
For effectively attacking the classifiers, the adversary should tune the hyperparameters required by the attack algorithms such as the number of iteration, the perturbation limitation, the number of line search within the manifold, which we express them all as the budget allocations. For finding the optimal required budgets, we bind the fooling rates of the attack algorithms to a predefined threshold $AUC > 0.9$ associated with the area under curve of the attack success. In other words, we allocate as much budget as needed for reaching the $AUC > 0.9$ for all attacks against the victim models. This is a critical threshold for demonstrating the extreme vulnerability of neural networks against adversarial attacks.

In accordance to the above note, we use Foolbox \cite{rauber2017foolbox}, the freely available python package in support of the uniform reproducible implementations of the attack algorithms. For the BIM-a and BIM-b algorithms, we define the $\epsilon \geq 0.0015$ with the confidence of ($\geq 75\%$). In the JSMA framework, we set the number of iterations to a maximum of 1000 and the scaling factor within $[0, 250]$ (with equivalent displacement of 50). The number of iterations in the DeepFool attack is initialized to 100 with the supremum value in light of 600 and the static step of 100. For the costly CWA attack, we set the search step $c \in \left \{1, 3, 5, 7 \right \}$ within the number of iteration $\left \{25, 100, 1\mathrm{k}, 1.5\mathrm{k} \right \}$ associated with every $c$. Except of the DeepFool which is a non-targeted attack, we randomly select targeted wrong labels for the rest of the algorithms.

There are four hyperparameters required for the black-box PIA algorithm. We empirically limit the perturbation bound to $\epsilon \geq 0.001$ followed by an iterative line search to find the most approximately optimal variance in the NES gradient estimation. We initialize the number of iteration to 500 with decay rate of $0.001$ and the learning rate $\eta \in [0.001, 0.6]$.

In the framework which we attack the front-end audio classifiers, we run the algorithms on the shuffled batches of 500 samples up to 50 batches of 100 samples randomly selected from the clean spectrograms in every step toward spanning the entire datasets. These attacks are performed considering the abovementioned allocated budgets once before and after adversarially training in order to measure the robustness of the models. Section \ref{sec:advtra} provides details on how adversarially training has been implemented.

\section{Adversarially Training}
\label{sec:advtra}
The idea of adversarially training was firstly proposed in \cite{goodfellow2014explaining}, where authors showed that, augmenting the training dataset with the one-shot FGSM adversarial examples improves the robustness of the victim models. As commonly known, the main advantage of this simple approach is that, it does not shatter nor obfuscate gradient information while runs a fast non-iterative procedure. This has made the adversarially training to be a relatively reliable defense approach. However, it may not confidently defend against stronger white-box adversarial algorithms \cite{tramer2017ensemble}.

Many adversarial defense approaches have been introduced during the past years which have been reported to outperform FGSM-based adversarially training \cite{papernot2016distillation, buckman2018thermometer, guo2017countering}. However, some studies have been reported that these advanced defense approaches shatter gradient vectors and they might easily break against strong adversarial attacks which do not incorporate the exact gradient information such as the backward pass differentiable approximation \cite{athalye2018obfuscated}.

Augmenting the clean training dataset with adversarial examples in the adversarially trained framework is shown in Eq.~\ref{eq:adv_tra} \cite{goodfellow2014explaining}.
\begin{equation}
    J{}'(\bold{x}, l, \bold{w})=\alpha J(\bold{x}, l, \bold{w})+(1-\alpha)J(\bold{x{}'}, l, \bold{w})
    \label{eq:adv_tra}
\end{equation}
\noindent where $\alpha$ is a subjective weight scalar definable by the adversary. Additionally, $J$ and $\bold{w}$ denote the loss function and the derived weight vector of the victim model, respectively. Moreover $\bold{x}$ and $\bold{x{}'}$ refer to the legitimate and adversarial example associated with the genuine label $l$. Adversarially training using a costly attack algorithm is very time-consuming and memory prohibitive in practice. Therefore, we use the FGSM for augmenting the original spectrogram datasets with the adversarial examples according to the assumption of $J{}'(\bold{x}, l, \bold{w})=J(\bold{x{}'}, l, \bold{w})$.

In the next section, we report our achieved results for the dense neural network models about the adversarial attacks and adversarially training on four different representations, namely STFT, DWT, STFT appended with tonnetz features, and DWT appended with tonnetz chromagrams.
\begin{table*}[htpb!]
\centering
\caption{Recognition performance (\%) of the audio classifiers trained on the original spectrogram datasets (without adversarial example augmentation). Values inside of the parenthesis indicate the recognition percentage drop after adversarially training the models with the fooling rate $AUC>0.9$. Accordingly, the maximum perturbation is achieved at $\left \| \epsilon \right \|_{2} \leq 3$. Outperforming accuracies are shown in bold face.}
\begin{tabular}{|c||c||c|c|c|c|c|c|}
\hline
Dataset                       & Representations & GoogLeNet          & AlexNet            & ResNet-18          & ResNet-34          & ResNet-56          & VGG-16             \\ \hline  \hline
\multirow{4}{*}{ESC-50}       & STFT            & $67.83$, $(06.89)$  & $64.32$, $(10.91)$ & $66.85$, $(12.13)$ & $67.21$, $(14.43)$ & $\textbf{69.77}$, $(09.29)$  & $68.94$, $(08.32)$  \\ \cline{2-8} 
                              & DWT             & $70.42$, $(08.42)$  & $65.39$, $(11.23)$ & $67.06$, $(15.71)$ & $67.55$, $(18.76)$ & $\textbf{71.56}$, $(11.09)$ & $71.43$, $(16.28)$ \\ \cline{2-8} 
                              & STFT $|$ Tonnetz  & $70.11$, $(24.09)$ & $64.21$, $(23.76)$ & $67.62$, $(19.48)$ & $66.75$, $(23.31)$ & $\textbf{70.22}$, $(25.19)$ & $70.18$, $(23.68)$ \\ \cline{2-8} 
                              & DWT $|$ Tonnetz   & $68.76$, $(19.07)$ & $68.31$, $(18.53)$ & $68.49$, $(24.27)$ & $67.15$, $(21.56)$ & $\textbf{71.79}$, $(18.21)$ & $68.37$, $(18.73)$ \\ \hline \hline
\multirow{4}{*}{UrbanSound8K} & STFT            & $88.32$, $(10.35)$ & $86.07$, $(21.43)$ & $88.24$, $(14.94)$ & $88.61$, $(09.19)$  & $\textbf{88.77}$, $(23.06)$ & $87.93$, $(14.66)$ \\ \cline{2-8} 
                              & DWT             & $90.10$, $(16.35)$ & $87.51$, $(19.59)$ & $88.07$, $(15.08)$ & $88.38$, $(19.04)$ & $\textbf{90.14}$, $(15.49)$ & $90.11$, $(16.35)$ \\ \cline{2-8} 
                              & STFT $|$ Tonnetz  & $88.44$, $(25.77)$ & $86.81$, $(22.05)$ & $88.13$, $(17.64)$ & $88.38$, $(26.42)$ & $89.41$, $(20.73)$ & $\textbf{89.42}$, $(21.38)$ \\ \cline{2-8} 
                              & DWT $|$ Tonnetz   & $89.32$, $(16.83)$ & $87.34$, $(20.41)$ & $88.76$, $(29.12)$ & $89.80$, $(27.45)$ & $\textbf{91.36}$, $(26.08)$ & $89.97$, $(24.56)$ \\ \hline
\end{tabular}
\label{table:recog}
\end{table*}
\begin{table*}[htpb!]
\centering
\caption{Robustness comparison (average $AUC\%$) of the adversarially trained models attacked with the constraint $\left \| \epsilon \right \|_{2} \leq 3$. Victim models with lower fooling rates are indicated in bold.}
\begin{tabular}{|c||c||c|c|c|c|c|c|}
\hline
Dataset                       & Representations  & GoogLeNet & AlexNet          & ResNet-18 & ResNet-34 & ResNet-56 & VGG-16  \\ \hline \hline
\multirow{4}{*}{ESC-50}       & STFT             & $53.12$   & $\textbf{50.97}$ & $51.13$   & $55.31$   & $53.87$   & $51.05$ \\ \cline{2-8} 
                              & DWT              & $55.68$   & $\textbf{51.03}$ & $52.56$   & $54.18$   & $52.26$   & $52.23$ \\ \cline{2-8} 
                              & STFT $|$ Tonnetz & $56.18$   & $\textbf{50.46}$ & $53.10$   & $55.29$   & $54.19$   & $52.82$ \\ \cline{2-8} 
                              & DWT $|$ Tonnetz  & $55.74$   & $\textbf{49.33}$ & $54.87$   & $53.77$   & $50.42$   & $51.37$ \\ \hline \hline
\multirow{4}{*}{UrbanSound8K} & STFT             & $56.09$   & $\textbf{53.24}$ & $54.08$   & $55.91$   & $57.30$   & $54.35$ \\ \cline{2-8} 
                              & DWT              & $58.98$   & $\textbf{51.92}$ & $53.59$   & $54.40$   & $55.86$   & $53.66$ \\ \cline{2-8} 
                              & STFT $|$ Tonnetz & $55.80$   & $\textbf{50.71}$ & $52.75$   & $51.02$   & $54.11$   & $52.39$ \\ \cline{2-8} 
                              & DWT $|$ Tonnetz  & $58.46$   & $\textbf{52.23}$ & $55.13$   & $56.81$   & $55.38$   & $55.26$ \\ \hline
\end{tabular}
\label{table:robust}
\end{table*}
\begin{table*}[htpb!]
\centering
\caption{Comparison of $\epsilon_{r}$ for attacking the original and adversarially trained models with the constraint of $AUC>0.9$. Higher values for $\epsilon_{r}$ associated with each representation are shown in bold.}
\begin{tabular}{|c||c||c|c|c|c|c|c|}
\hline
Dataset                       & Representations  & GoogLeNet & AlexNet & ResNet-18 & ResNet-34 & ResNet-56 & VGG-16  \\ \hline \hline
\multirow{4}{*}{ESC-50}       & STFT             & $1.412$   & $1.631$ & $1.897$   & $2.154$   & $\textbf{2.312}$   & $2.107$ \\ \cline{2-8} 
                              & DWT              & $1.562$   & $1.509$ & $1.741$   & $1.982$   & $1.976$   & $\textbf{2.307}$ \\ \cline{2-8} 
                              & STFT $|$ Tonnetz & $1.804$   & $1.918$ & $2.003$   & $\textbf{2.161}$   & $2.095$   & $1.674$ \\ \cline{2-8} 
                              & DWT $|$ Tonnetz  & $2.014$   & $2.336$ & $1.788$   & $1.903$   & $\textbf{2.609}$   & $2.230$ \\ \hline \hline
\multirow{4}{*}{UrbanSound8K} & STFT             & $1.562$   & $1.903$ & $\textbf{2.439}$   & $1.372$   & $1.991$   & $1.703$ \\ \cline{2-8} 
                              & DWT              & $2.154$   & $2.287$ & $2.764$   & $1.644$   & $\textbf{2.892}$   & $1.789$ \\ \cline{2-8} 
                              & STFT $|$ Tonnetz & $2.231$   & $2.108$ & $1.981$   & $2.003$   & $1.401$   & $\textbf{2.308}$ \\ \cline{2-8} 
                              & DWT $|$ Tonnetz  & $1.606$   & $2.199$ & $2.405$   & $1.604$   & $\textbf{2.501}$   & $1.702$ \\ \hline
\end{tabular}
\label{table:epsi}
\end{table*}

\section{Experimental Results}
\label{sec:expersut}
We conduct our experiments on two environmental sounds datasets: UrabanSound8K \cite{Salamon:UrbanSound:ACMMM:14} and ESC-50 \cite{piczak2015esc}. The first dataset contains 8732 short recording arranged in 10 classes (car horn, dog bark, drilling, jackhammer, street music, siren, children playing, air conditioner, engine idling and gun shot) with the audio length of $<4$ seconds. ESC-50 dataset contains 2K audio signals with an equal length of five seconds organized in 50 classes.

For enhancing both quality and quantity of these datasets, especially for ESC-50, we filter samples using the pitch-shifting operation in the temporal domain as proposed in \cite{esmaeilpour2020unsupervised}. According to their proposed 1D filtration setup, we use the scales of $\left \{ 0.75, 0.9, 1.15, 1.5 \right \}$. This increases the size of the datasets by a factor of 4. 

Following the explanations provided in section \ref{sec:specgen} about the spectrogram production, the dimension of each resulting spectrogram is $1568 \times 768$ for both STFT and DWT (the logarithmic scale) representations on the two datasets. Moreover, the dimensions of the resulting chromagrams is $1568 \times 540$, which will be appended to the aforementioned representations. Table~\ref{table:recog} summarizes recognition accuracies of the classifiers trained on these spectrograms. Additionally, this table shows the effect of the adversarially training on the recognition performance of these models. 

The classifiers in Table~\ref{table:recog} have been selected for evaluation on the test sets after running the five-fold cross-validation scenario on the randomized development portion of the training datasets. Regarding this table, different architectures of the deep neural networks show competitive performances. However, in the majority of the cases, the ResNet-56 outperforms other classifiers averaged over 10 repeated experiments on the spectrograms. The highest recognition accuracy has been achieved by the ResNet-56 architecture,  trained on the appended representation of DWT and tonnetz chromagrams for both UrbanSound8K and ESC-50 datasets. The number of parameters in the ResNet-56 is 11.3\% and 14.26\% higher than its rival models VGG-16 and ResNet-34, respectively.

Fig.~\ref{attack_spect} visually compares the adversarial examples crafted against the outperforming classifier, the ResNet-56, using the six adversarial attacks with a randomly selected audio sample and represented with the four spectrograms approaches described earlier. Although the generated spectrograms are visually very similar to their legitimate counterparts, they all make the classifier to predict wrong labels.

Table~\ref{table:recog} also shows the drop ratio of the recognition accuracies after adversarially trained the models following the procedure explained in section~\ref{sec:advtra}. The maximum required adversarial perturbation for complying with the fooling rate of $AUC>0.9$ is achieved at $\left \| \epsilon \right \|_{2} \leq 3$, averaged over all the attacks. In attacking the adversarially trained models, the procedures outlined in section~\ref{adv:setup} has been implemented individually for every audio classifier. According to the obtained results,  adversarially training considerably reduces the performance of all models. For the ESC-50, the neural networks trained on the appended representation of STFT and tonnetz features (STFT $|$ Tonnetz) has experienced the most negative impact compared to other representations. The average drop ratio for adversarially trained models on the DWT $|$ Tonnetz representations is slightly more than the STFT $|$ Tonnetz counterparts for the UrbanSound8K dataset. However, for both datasets, these ratio for models trained on the DWT spectrogram are considerably higher than those trained with the STFT representations. 

We measure the fooling rate of adversarially trained models after attacking them using the same six adversarial algorithms following the procedure explained in section~\ref{adv:setup} with the imposed condition of $\left \| \epsilon \right \|_{2} \leq 3$ for the adversarial perturbation. This experiment uncovers the impact of adversarially training on the robustness of the audio classifiers (see Table~\ref{table:robust}). We applied the aforementioned condition to make this table comparable with Table~\ref{table:recog}. Regarding the results reported in Table~\ref{table:robust}, adversarially training has improved the robustness of all the classifiers, particularly AlexNet. 

For investigating the overall impact of the adversarially training on the robustness of audio classifiers, we attack the adversarially trained models using the same six attack algorithms without the condition of $\left \| \epsilon \right \|_{2} \leq 3$. Unfortunately, we could achieve the fooling rate with $AUC>0.9$ for all the classifiers following the attack procedure explained in section~\ref{adv:setup}. However, attacking the adversarially trained models requires larger values for the adversarial perturbation ($\left \| \epsilon \right \|_{2}$) compared to attacking the original models and consequently, increases the number of callbacks to the original spectrogram with extra batch gradient computations. This might degrade the quality of the generated spectrograms. In order to analytically compare the maximum adversarial perturbation required for the original and the adversarially trained models, we compute the average perturbation ratio as shown in Eq.~\ref{eq:epsilonratio}:
\begin{equation}
    \epsilon_{r}=\left | \frac{\epsilon_{a}}{\epsilon_{o}} \right |
    \label{eq:epsilonratio}
\end{equation}
\noindent where $\epsilon_{a}$ and $\epsilon_{o}$ denote the average adversarial perturbation required for successfully attacking the adversarially trained and original models (both with $AUC>0.9$), respectively. Table~\ref{table:epsi} summarizes values for $\epsilon_{r}$ for the victim models trained on different representations.

Note that an $\epsilon_{r} \geq 1$ indicates the positive impact of adversarially training on the robustness of the audio classifiers via increasing the computational cost of the attack by expanding the magnitude of the required adversarial perturbation. With respect to the measured $\epsilon_{r}$ metric for all the front-end classifiers, the ResNet-56 architecture showed better robustness against adversarial attacks in average for 50\% of the experiments. In other words, attacking this model adds additional cost for the adversary in crafting adversarial examples with the $AUC > 0.9$.

\section{Conclusion}
In this paper, we presented the impact of adversarially training as a gradient obfuscation-free defense approach against adversarial attacks. We trained six advanced deep learning classifiers on four different 2D representations of environmental audio signals and run five white-box and one black-box attack algorithms against these victim models. We demonstrated that adversarially training considerably reduces the recognition accuracy of the classifier but improves the robustness against six types of targeted and non-targeted adversarial examples by constraining over the maximum required adversarial perturbation to $\left \| \epsilon \right \|_{2} \leq 3$. In other words, adversarially training is not a remedy for the threat of adversarial attacks, however it escalates the cost of attack for the adversary with demanding larger adversarial perturbations compared to the non-adversarially trained models.  

\section*{Acknowledgment}
This work was funded by the Natural Sciences and Engineering Research Council of Canada (NSERC) under Grant RGPIN 2016-04855 and Grant RGPIN 2016-06628.

\bibliographystyle{IEEEtran}
\bibliography{mybib}

\end{document}